\documentclass[%
 aip,
 amsmath,amssymb,
 reprint,%
]{revtex4-1}

\usepackage{graphicx}
\usepackage{dcolumn}
\usepackage{bm}

\usepackage[utf8]{inputenc}
\usepackage[T1]{fontenc}
\usepackage{mathptmx}
\usepackage{etoolbox}

\makeatletter
\def\@email#1#2{%
 \endgroup
 \patchcmd{\titleblock@produce}
  {\frontmatter@RRAPformat}
  {\frontmatter@RRAPformat{\produce@RRAP{*#1\href{mailto:#2}{#2}}}\frontmatter@RRAPformat}
  {}{}
}%
\makeatother
\begin{document}

\preprint{AIP/123-QED}

\title{Photon-photon coupling induced bound state in the 
continuum and transparency}

\author{Ekta Tunwal}
\altaffiliation[These authors]{ contributed equally to this work.}
 \affiliation{Department of physics, Indian Institute of Technology (Banaras Hindu University), Varanasi-221005, BHARAT (India)}
 
\author{Kuldeep Kumar Shrivastava}
\altaffiliation[These authors]{ contributed equally to this work.}
 \email{kuldeep224@gmail.com, biswanath.phy@iitbhu.ac.in}
\affiliation{Department of physics, Indian Institute of Technology (Banaras Hindu University), Varanasi-221005, BHARAT (India)}
\author{Rakesh Kumar Nayak}
\affiliation{Department of physics, Indian Institute of Technology (Banaras Hindu University), Varanasi-221005, BHARAT (India)}
\author{Ravi Kumar}
\affiliation{Department of Electronics Engineering, Indian Institute of Technology (Banaras Hindu University), Varanasi-221005, BHARAT (India)}
\author{Somak Bhattacharyya}
\affiliation{Department of Electronics Engineering, Indian Institute of Technology (Banaras Hindu University), Varanasi-221005, BHARAT (India)}
\author{Rajeev Singh}
\affiliation{Department of physics, Indian Institute of Technology (Banaras Hindu University), Varanasi-221005, BHARAT (India)}
\author{Biswanath Bhoi}
\affiliation{Department of physics, Indian Institute of Technology (Banaras Hindu University), Varanasi-221005, BHARAT (India)}

\date{\today}

\begin{abstract}
This study presents the coherent and dissipative coupling realized in the hybrid photonic resonators that have been achieved via the constructive and destructive interference of the photonic resonator fields with the radiation of a common transmission line fed with microwave photons. In the dissipative coupling regime we have found the coexistence of a peculiar phenomenon bound state in the continuum (BIC) near the crossing of frequency of the uncoupled resonators by satisfying the Friedrich-Wintgen BICs condition. Again just by rotating one of the samples and with the dynamic adjustment of a parameter we have achieved coupling induced transparency between the photonic resonators. This transition from BIC in the absorption regime to transparency opens avenues for different sorts of plain or programmable oscillators, filters, quantum information processors, sensors etc.
 
\end{abstract}

\maketitle

\section{INTRODUCTION}
The modern era of science and technology is doubtlessly wave-based and is central to physics and engineering both. This era requires full prospects of light matter interactions, their control and interplay to enhance their potential application in cavity quantum devices, quantum materials, oscillators, filters, sensors, waveguide, switches, diverse quantum technologies, etc \cite{wei2022towards,hsu2016bound,pozar2021microwave,tiwari2024modified}.
While the unveiling of light matter interaction is going on and seems to be an endless endeavor, some of the key and peculiar phenomena already significant are electromagnetically induced absorption (EIA), electromagnetically induced transparency (EIT), bound state in the continuum (BIC), etc\cite{hsu2016bound,yu2022observation,rao2021controlling,bhoi2022coupling,moiseyev2011non,shrivastava2024emergence}.

EIA, which is a consequence of greater absorption of transmission signals having a key role in developing new avenues for controlling and utilizing light matter interactions, which are very much less explored until now. Analogous of the EIA phenomenon, in the linear domain is coupling induced absorption (CIA) that is nowadays drawing much attention due to dissipative coupling in the form of CIA/ EIA arises when there is a significant dissipation of energy at some frequency. That is crucial for optical isolators, nonreciprocal devices for better sensing techniques etc\cite{bhoi2022coupling,shrivastava2024emergence}.

Another phenomenon that is relatively more explored is EIT, which is a consequence of reduction in resonance absorption  because of the medium becoming transparent near/ around the coupling center. Also analogous to EIT, in the linear domain is coupling induced transparency (CIT). This well known form of light-matter interaction comes from the phase relationship between different modes of the system in which coherent exchange of energy is happening between different subsystems of the system\cite{bhoi2022coupling,tiwari2024modified}.  

In 1929, BICs were first proposed in quantum mechanics, but they are peculiar wave phenomena that have since been identified in electromagnetic, water, elastic and acoustic waves and in different material systems etc. BIC, are peculiar type of waves that remain confined and coexist with a continuous spectrum of radiating waves that may carry energy away. The BIC results when two modes of a system become dissipative in nature and their coupling results in a destructive interference. They are essential for different piezoelectric materials, fibers, optical waveguides, topological insulators, waveguides, novel quantum information processors, etc.\cite{hsu2016bound,yu2022observation,rao2021controlling}.

More recently, various systems reported BIC in engineered quantum potential\cite{von1929uber}, coupled waveguide array\cite{dreisow2009adiabatic}, system protected acoustic waveguides\cite{plotnik2011experimental},  periodic metal grid\cite{ulrich1975symposium}, distributed feedback laser with 1D periodicity\cite{henry1985observation}, a quantum well based on separability\cite{robnik1986simple}, water waves by engineering the shapes of two obstacles\cite{mciver1996example}, acoustic waves\cite{linton2007embedded}, two interfering resonances\cite{friedrich1985interfering}, single resonance in a slab\cite{hsu2013observation}, topological properties in a slab\cite{zhen2014topological}, surface of a piezoelectric solid\cite{yamanouchi1972propagation}, parity time symmetric system\cite{longhi2014bound}, tight binding lattice with a engineering hoping rates\cite{corrielli2013observation}, high output power having low threshold at room temperature\cite{hirose2014watt}, asymmetric coupled fiber loops\cite{regensburger2013observation} etc. But there is always a keen interest in miniaturizing or simplifying the device\cite{hsu2016bound,verma2021introduction,baraclough2018investigation}. 

In our work, we have developed a planar device capable of holding not only BIC, but CIT and CIA phenomena simultaneously. We have explained these phenomena within a microwave photon-photon coupled system. A notable achievement of our work is that we have achieved BIC and CIT using only linear coupling. We have also shown the result obtained from simulation, theory as well as experiments by exciting the photon modes at room temperature. Finally in a single device we achieve fine control for BIC to happen and later with some adjustment have shown switching BIC to CIT. Have explained the results using a quantum model and simple planar resonators suggesting a way to observe and control these phenomena in similar simple and suitable devices which may have various technological applications.

\section{MODEL}

Different coupled modes of a hybrid system connected to a channel additionally affect each other through the channel, which can be in the form  of bath,\ cavity,\ microstripline etc., to inject travelling photons in the system and is  also attached to input and output ports. We may write a general Hamiltonian for such a system as \cite{scully1997quantum,walls2008quantum}

\begin{align} \label{E1}
    H / \hbar &= \omega_1\hat X_1 ^\dag \hat X_1 + \omega_2\hat X_2 ^\dag \hat X_2 + \Delta(\hat X_1 +\hat X_1^\dag)(\hat X_2 +\hat X_2^\dag) \nonumber \\ &\quad +\int\omega_k \hat p_k^\dag \hat p_k dk \nonumber \\ &\quad +\int\left[\{\lambda_1 (\hat X_1 +\hat X_1^\dag)+\lambda_2 (\hat X_2 +\hat X_2^\dag)\}(\hat p_k+\hat p_k^{\dag})\right] dk.
 \end{align} 
where $\omega_1$ and $\omega_2$ are the resonance frequencies of the two modes. The first integral term of the Hamiltonian is for the feeding channel 
(cavity\ bath\ microstripline etc. depending upon the case) connected to the input and output ports. In our Hamiltonian formulation, traveling photons mediate the interaction while passing through the feeding channel, which we are integrating over a real domain from $-\infty $ to $+\infty $. Travelling photons are bosonic in nature and its creation (annihilation) operator is denoted by ${\hat p_k}^\dag (\hat p_k)$ which obeys
$[\hat p_k, \hat p_{k'}^\dag] = \delta(k-k')$.

Here $k$ represents the wave vector, where $\omega_k$ denotes the frequency of travelling photons. The last term accounts for the interaction between the traveling photons and each mode, the interaction strength between them is denoted by $\lambda_l$, linearly modelled in ${\hat p_k}^\dag (\hat p_k)$.

After taking rotating wave approximation (RWA) we may write Eq.~\ref{E1} as\cite{moiseyev2011non,tiwari2024modified}

\begin{align} \label{E2}
    H/ \hbar &=\omega_1 \hat X_1^\dag \hat X_1 + \omega_2 \hat X_2^\dag \hat X_2 +
    \int\omega_k \hat p_k^\dag \hat p_k dk +\Delta(\hat X_1 \hat X_2^\dag + \hat X_2 \hat X_1^\dag) \nonumber \\
    & + \int\left[\lambda_1 (\hat X_1 \hat p_k^\dag +\hat X_1^\dag \hat p_k)+\lambda_2 (\hat X_2 \hat p_k^\dag +\hat X_2^\dag\hat p_k)\right]dk , 
 \end{align} 
 We have modelled the two mode coupled photon-photon hybrid quantum system  using the Hamiltonian of Eq.~\ref{E2} having 1st photon mode($P_1$) and 2nd photon modes ($P_2$).

\begin{figure}
    \centering
    \includegraphics[width=\linewidth]{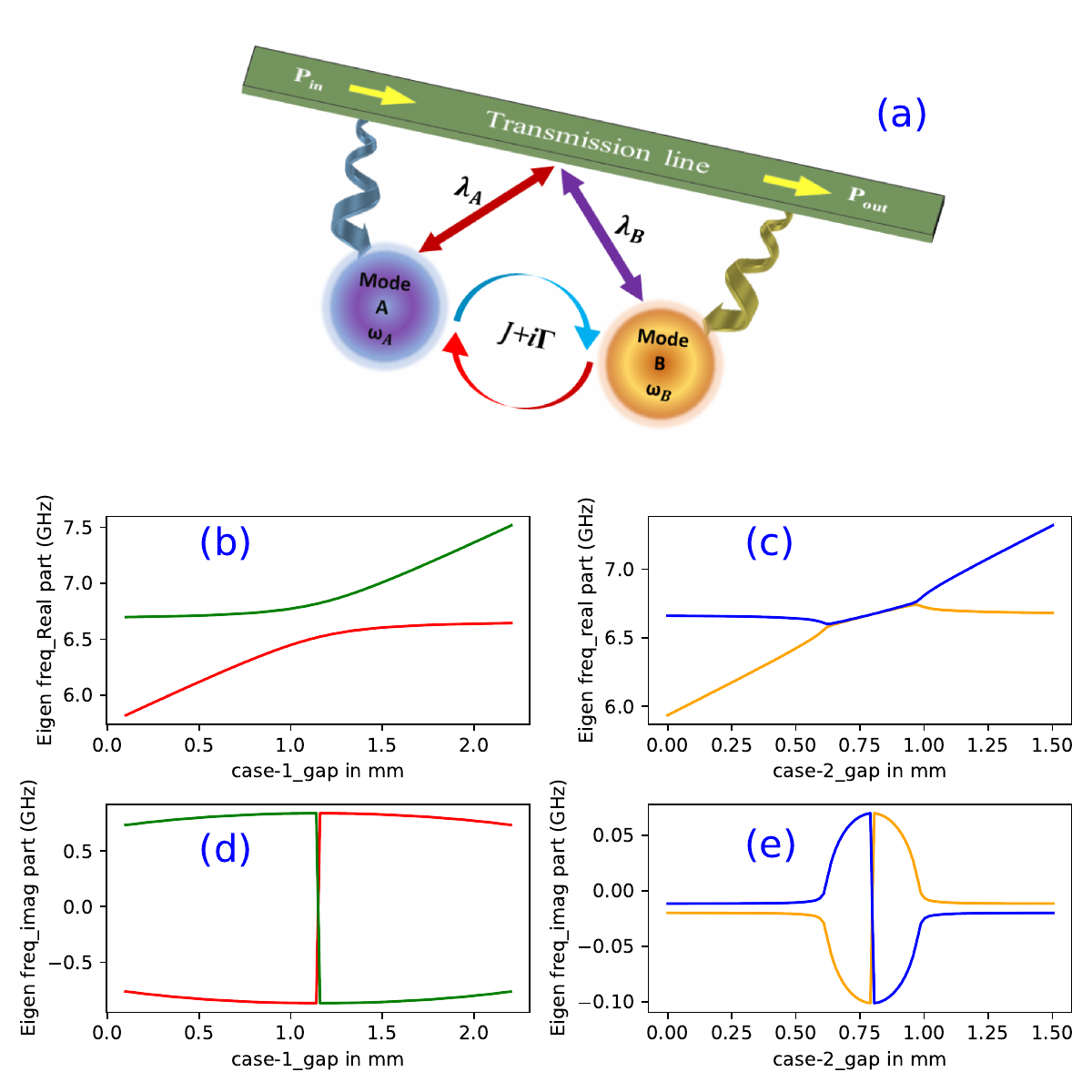}
    \caption{(a) Schematic representation of two modes interactions in a coupled system. (b/ d) showing CIT/ level repulsion (LR) due to coherent coupling representing the real/ imaginary part of eigenvalues. (c/ e) showing CIA/ level attraction (LA) due to dissipative coupling representing the real/ imaginary part of eigenvalues. }
    \label{F1}
\end{figure}
 
 It is well established now that EIA/ CIA/ LA or EIT/ CIT/ LR is determined by the combined and relative strengths and phases of the oscillating magnetic fields generated from the resonator's split gaps and the travelling waves of MSL. Also the complex coupling constant ($\Delta= J+i\Gamma$) has a dominating real part ($J$) for coherent coupling shown in Fig.~\ref{F1} (b) and (d). While for dissipative coupling it has a dominating imaginary part ($\Gamma$) shown in Fig.~\ref{F1} (c) and (e). BIC is a special case of dissipative coupling that arises when the damping of the modes overcome the contribution of real part of complex coupling even with no imaginary coupling\cite{friedrich1985interfering,yu2022observation,hsu2016bound}. 
 
 for this system the Heisenberg-Langevin equations can be compactly written in a matrix form as\cite{shrivastava2024emergence, tiwari2024modified}
 \begin{align} \label{E3}
     \frac{d}{dt} \begin{bmatrix} \hat X_1\\ \hat X_2
     \end{bmatrix}
     = -i\,H_{coupling}
      \begin{bmatrix} \hat X_1\\ \hat X_2
     \end{bmatrix}
     -i\begin{bmatrix} \sqrt{\beta_1}\\\sqrt{\beta_2}
     \end{bmatrix} \hat P_{in} (t)
 \end{align} 
where, 
 \begin{center}
 $ H_{coupling}= \begin{bmatrix}
    &\omega_1-i\beta_1
    &\Delta-i\sqrt{\beta_1 \beta_2}\\
    &\Delta-i\sqrt{\beta_1 \beta_2}
    &\omega_2-i\beta_2
    
     \end{bmatrix}$    
 \end{center} 
     
 \noindent where $\beta_1$, $\beta_2$ are damping rates for the $P_1$ and $P_2$ modes respectively.

 Whenever the  direct interaction between hybrid photon modes happens, the resonance frequencies of the hybridized modes are splitted into lower and higher branches for LR but for LA they get merged around coupling center , that can be analytically solved by the use of coupling matrix ($H_{coupling}$) of the Eq.~\ref{E3} for real parts of its eigenvalues. 

 The condition for a BIC for coupled resonators can be obtained from $H_{coupling}$ as
\begin{align} \label{E4}
 \Delta (\beta_1 - \beta_2)= \sqrt{\beta_1 \beta_2}(\omega_1 - \omega_2)   
\end{align}
and is known as the Friedrich-Wintgen condition\cite{friedrich1985interfering}.
In this system the two resonators radiate into the same channel, so the continuum coupling term $\sqrt{\beta_1 \beta_2}$ is coming out of result of interference of radiation. Here one of the eigenvalues is becoming more lossy while one of the eigenvalue becomes purely real and turns into a BIC. It is obvious that when $\Delta$=0 or when $\beta_1 = \beta_2$, the BIC is obtained at $\omega_1 = \omega_2$. Therefore, when $\Delta \approx 0$ or  $\beta_1 \approx \beta_2$, Friedrich- Wintgen BICs are found near the crossing of frequency of uncoupled resonators \cite{hsu2016bound}. Generally when the number of resonances surpasses the number of radiation channels, these BICs are possible \cite{remacle1990trapping,berkovits2004discrete}.

 Eq.~\ref{E3} is in the time forward domain having $\hat P_{in} (t)$, we can write it in the frequency domain and solve it for $\hat P_{in} (\omega)$. 
 Similarly, equivalent to Eq.~\ref{E3}, we can write another set of equations in time retarded domain having $\hat P_{out} (t)$, further we can write it in the frequency domain and solve it for  for $\hat P_{out} (\omega)$. Now we can write the relation between $P_{out}$ and $P_{in}$ in the frequency domain as \cite{shrivastava2024emergence}
\begin{align} \label{E5}
\hat P_{out}(\omega)-\hat P_{in}(\omega)= - 2i \left[\sqrt{\beta_1} \hat{X}_1 (\omega) + \sqrt{\beta_2} \hat X_2 (\omega) \right] 
\end{align}

For calculating the dispersion spectra of different types of interactions by taking into account all the cooperative effect of both photon modes and traveling photon modes of the combined hybrid system first we require to calculate $S_{21}$ (on $\frac{\omega}{2\pi} - H_{dc}$ plane), for two photon modes this can be written as
\begin{align} \label{E6}
S_{21}=\frac{\hat P_{out}}{\hat P_{in}}-1= - \frac{2i}{\hat P_{in}} \left[\sqrt{\beta_1} \hat{X}_1 + \sqrt{\beta_2} \hat X_2 \right] 
\end{align}

\section{Design of the hybrid system}

\begin{figure}
    \centering
    \includegraphics[width=\linewidth]{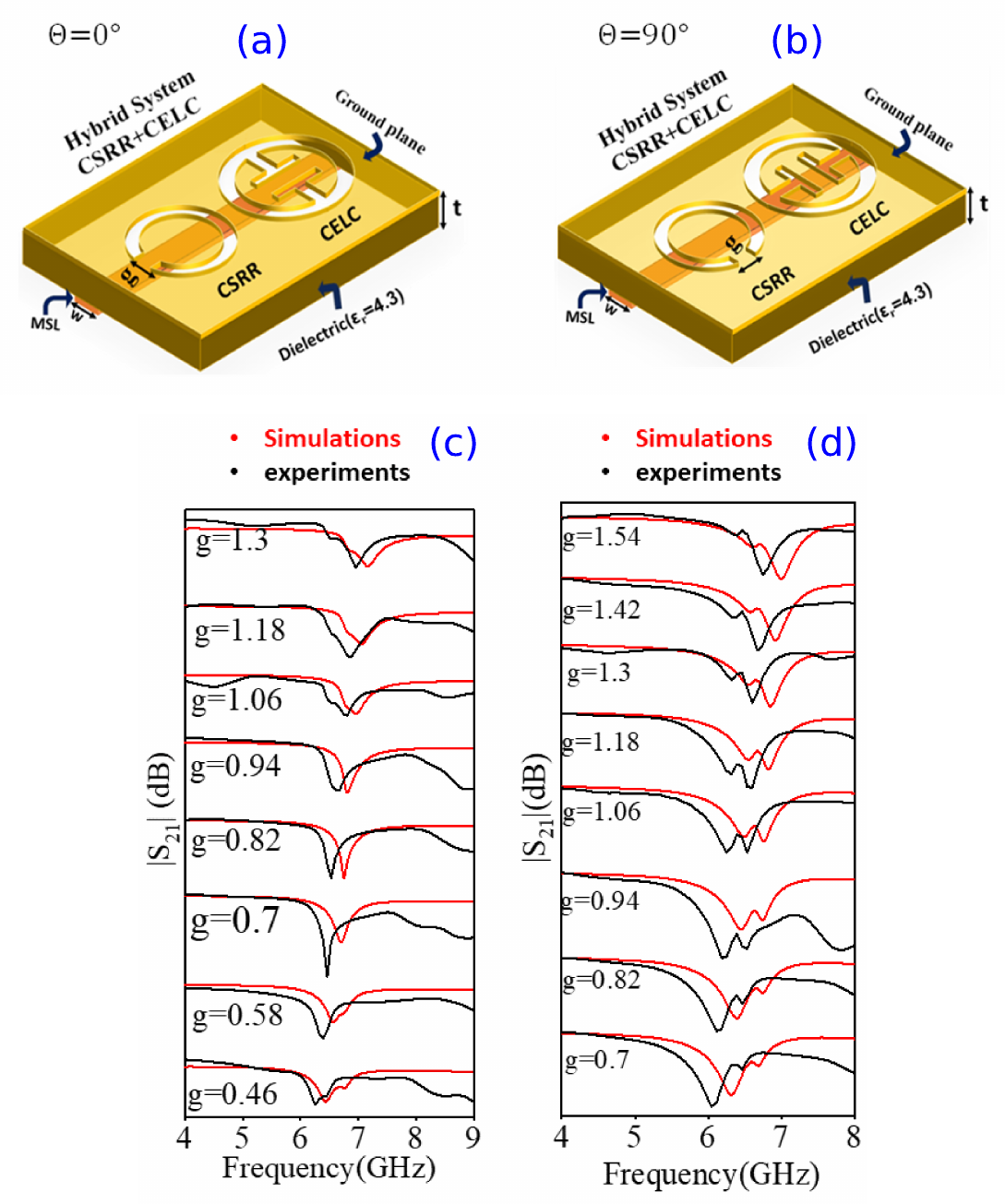}
    \caption{(a) and (b) Geometric 3D models for the simulation setup of the planar hybrid system for case 1 and case 2, respectively.(c) and (d) Demonstrate the consistency between simulated and experimental data, showing the varying nature of the transmission spectra $S_{21}$ plots through eigenfrequency values stack plots for both the cases, $g$ denotes the split gap of CSRR.}
    \label{F2}
\end{figure}

We have experimented with a planar hybrid system consisting of two photonic resonators, a Complementary Split-Ring Resonator (CSRR) and a Complementary Electric Inductive-Capacitive Resonator (CELC). This hybrid system leverages the strengths of both resonators, leading to enhanced performance and new possibilities for practical applications. The proposed design features a three-layer planar structure, as shown in Fig.~\ref{F2}(a) and (b). The base layer, acting as the ground plane, is made of copper with a thickness of 0.035 mm. Above this, the middle layer is a dielectric material. This substrate has dimensions of 40 mm by 40 mm by 1.6 mm and a relative permittivity of 4.3, providing a stable platform for the resonators. The CSRR and CELC are strategically placed on the ground plane of this substrate, allowing them to be excited by the currents induced by the transverse microwave magnetic field when microwave currents traverse the microstrip feeding line. Above the substrate, a microstrip line is engineered with a width of 3.064 mm, a length of 40 mm, and a thickness of 35~$\mu m$. This microstrip line, situated on the top surface of the substrate, serves as a waveguide, directing the flow of electromagnetic waves through the system. It is designed to achieve a characteristic impedance of 50~$\Omega$, aligning with the standard impedance of most RF systems. The configuration of the microstrip line is such that from the central strip electric fields originate and terminate perpendicularly to the ground plane\cite{verma2021introduction}.

The hybrid system has specific dimensions of CSRR, CELC, planer substrate and microstrip line. The Split ring resonators can be easily scaled in different size so as to enable operation over a wide range of frequencies, and they have been analysed from the low GHz up to visible frequencies \cite{baraclough2018investigation}.On the ground plane of this substrate, both resonators are symmetrically placed   about centre of ground plane, 0.1mm apart from each other.  For further understanding of the planar system, we conducted numerical simulations using CST Microwave Studio, a sophisticated electromagnetic full-wave solver. Our simulations focused on the electromagnetic field behavior within the frequency range of 1 to 12GHz. The specific dimensions of the CELC and CSRR and distance between the resonators has been optimized to ensure strong photon-photon coupling. We focused our simulations on understanding the impact of geometry and orientation of the resonators on the electromagnetic coupling of the CSRR and the CELC Resonator. To this end, we simulated our hybrid system under two distinct cases,

Case 1: In this scenario, we kept the geometric parameters of the CSRR constant, except for the split gap ‘g’, which varied systematically from 0.1 to 2.2 mm. This allowed us to observe the geometric effect on the coupling of the resonator. Meanwhile, the parameters for the CELC Resonator remained unchanged throughout the simulation process. This consistency ensured that any observed effects could be attributed to the variations in the CSRR’s split gap. The resonators were positioned such that the split gap of both the CSRR and CELC was perpendicular to the microstrip line orientation($\theta$=0°), shown in Fig.~\ref{F2}(a).

Case 2: To understand the impact of orientation on the coupling of the resonators, we rotated our system by 90 degrees ($\theta$=90°). This rotation aligned the split gap of both the CSRR and CELC parallel to the microstrip line orientation. Similar to Case 1, we varied the split gap ‘g’ of the CSRR from 0.1 to 2.2 mm while keeping the other geometric parameters constant, shown in Fig.~\ref{F2}(b).
By varying the split gap ‘g’ and observing the resulting changes in the system’s response by rotation of the system also, we were able to gain valuable insights into the design’s effectiveness and the resonators’ coupling behaviour.

\section{Results and Discussion}

This arrangement of the hybrid system gives rise to a microwave magnetic field encircling the microstrip line, ensuring an efficient transmission of electromagnetic waves. In this setup, the resonators operate as a parallel LC resonant circuit, leading to a quasi-static resonant effect\cite{pozar2021microwave}. This effect provides insight into the photon-photon interaction within the resonators. The specific dimensions of the resonators, particularly the split gaps, allow for precise control over their resonant frequencies and coupling strengths, enabling strong photon-photon coupling.

 The simulations were designed to observe the changes in the electric and magnetic fields of the resonators, which are influenced by factors such as orientation, position, geometry, and electromagnetic properties. These changes impact the destructive or constructive interference that governs the electromagnetic coupling. We simulated our planar system for the particular orientation of the split gap of resonators with respect to microstrip line alignment.

The transmission spectra observed for a specific resonator is a direct consequence of the excitation of photon modes within that resonator. This excitation, driven by the interaction of the resonator with the electromagnetic field, gives rise to distinct peaks in the transmission spectra, each corresponding to a specific resonant mode of the system.  Photon mode excitation in a CSRR and CELC resonator occurs when an alternating current of microwave frequency is transmitted through the microstrip line, it giving birth to a microwave magnetic field. A substantial part of this field permeates the CSRR and CELC, setting the stage for resonance. The CSRR and CELC, are designed to resonate at a specific frequency, which is determined by its physical dimensions and the properties of the material from which it is made. When the frequency of the microwave magnetic field matches the resonant frequency of the CSRR and CELC, the resonators absorb energy from the field. This absorbed energy sets the electrons within the CSRR and CELC into motion, causing them to oscillate. 

\begin{figure}
    \centering
    \includegraphics[width=\linewidth]{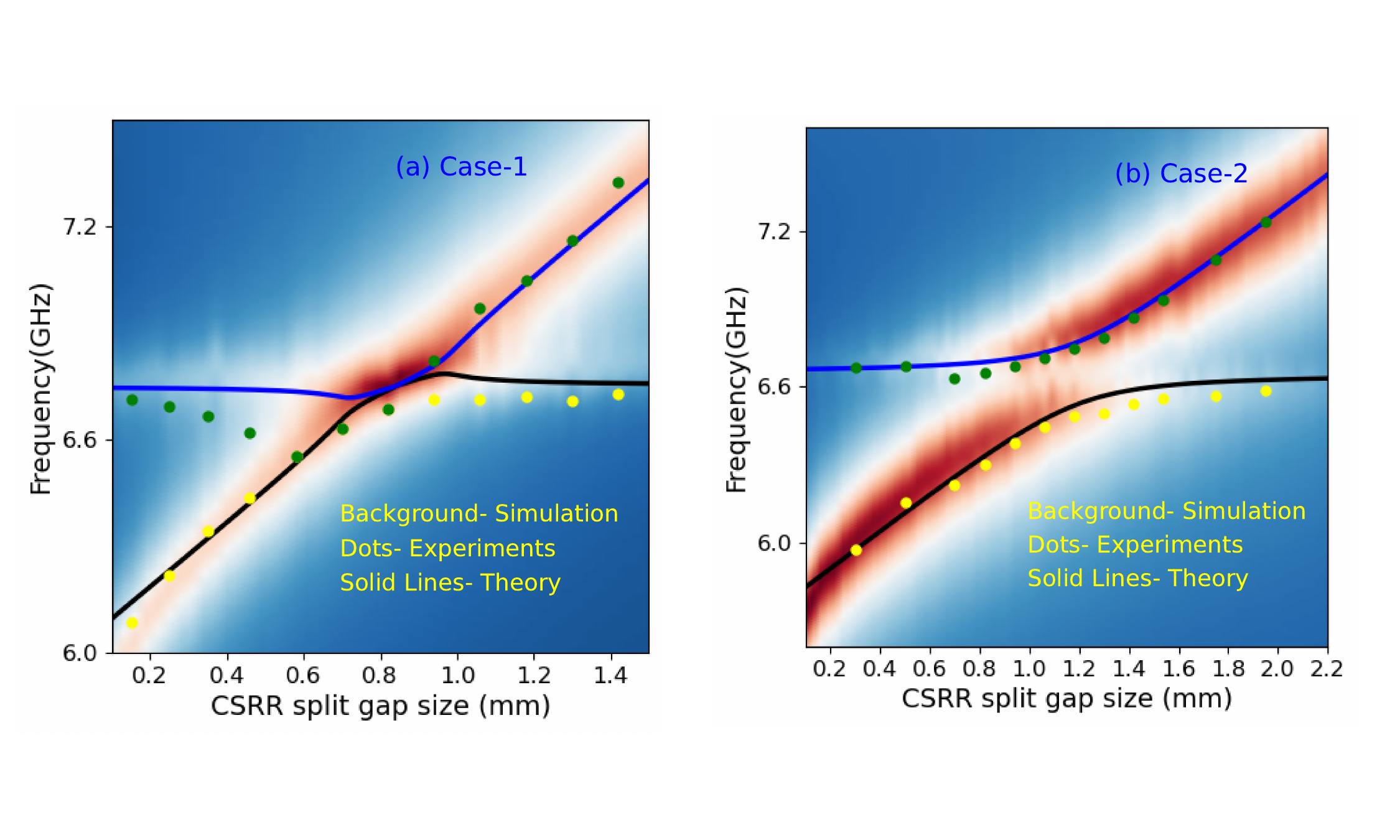}
    \caption{Colourmaps in the background represent the simulated transmission parameter $S_{21}$ for Case-1 and Case-2 respectively. Dots are taken from experimental measurements while Solid lines are drawn from theory. (a) Case-1: CSRR split gap size 'g' varies from 0.1~mm to 1.5~mm, at/ near the coupling center BIC is happening. (b) Case-2: CSRR split gap size 'g' varies from 0.1~mm to 2.2~mm, shows CIT. For both the cases the dimensions of CELC are kept constant.}
    \label{F3}
\end{figure}

In both the CSRR and CELC the electric and magnetic fields are intertwined and their distributions are closely related. These three distributions - electric field, magnetic field, and surface current - collectively influence the photon mode excitation and the resulting transmission spectra in the CSRR and CELC. 
Understanding these distributions provides valuable insights into the mechanism of these resonators. It will give intuitive understanding of the mechanism of the interaction between two resonators and how exchange of energy is happening.
The surface current distribution is a result of the interaction between the incident electromagnetic wave and the resonator. When the incident wave interacts with the resonator, it induces a current on the surface of the resonator. This induced current then generates a secondary electromagnetic wave, which interferes with the incident wave. The pattern of this interference determines the distribution of the surface current. The surface current   directly influences the strength and direction of the secondary wave. This, in turn, affects the overall electromagnetic field distribution within the resonator.

\begin{figure}
    \centering
    \includegraphics[width=\linewidth]{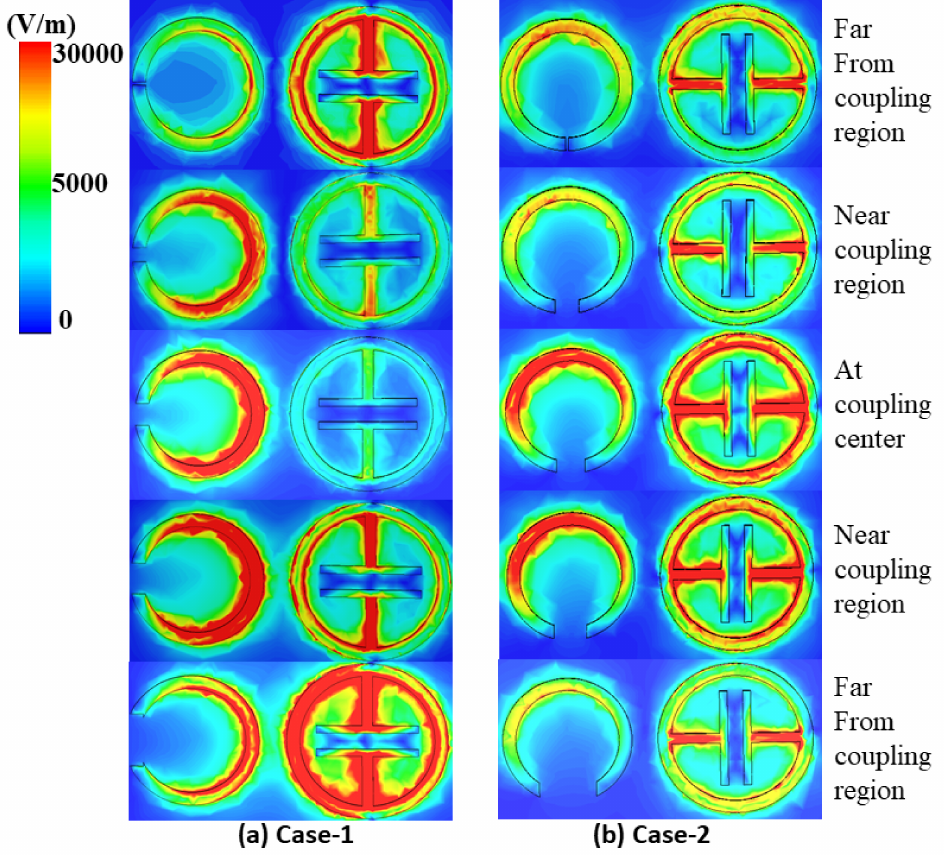}
    \caption{BIC, CIT and CIA also understood through field distribution study. The figures showcase the electric field distribution of a hybrid system for (a) Case-1, (b) Case-2. These regions are differentiated by varying the split gap of CSRR, for both the cases the dimensions of CELC are kept constant.}
    \label{F4}
\end{figure}

The field is concentrated around the gaps of the CSRR, where the surface currents are also dense. This is because the surface currents create the magnetic field, and the gaps in the CSRR structure focus the field intensity. For the CELC, the magnetic field shows a strong central concentration, aligning with the dense surface currents in the middle of the structure. Fig.~\ref{F4} showing electric field distribution for CSRR and CELC.  For the CSRR, the electric field is primarily concentrated around the ring structure rather than at the split gap. This concentration around the ring indicates where the capacitive effects are most significant. In the case of the CELC, the electric field is initially concentrated within the structure. This is because the CELC has a more complex geometry compared to the CSRR, which leads to a more intricate electric field distribution. The electric field lines tend to follow the path of least impedance, which is determined by the geometry of the resonators. Therefore, any change in the geometry or orientation of the resonators can lead to significant changes in the electric field distribution. 

Transmission parameters ($S_{21}$) have been recorded for the coupled CSRR-CELC hybrid system as a function of the microwave frequency flowing in the microstrip line. This was done for above-mentioned two cases with different values of the CSRR’s split gap, representing two distinct cases. Fig.~\ref{F3} illustrates the |$S_{21}$| spectra of the CSRR-CELC hybrid system. 

Satisfying the BIC condition as a peculiar outcome of CIA: for case-1, Fig.~\ref{F2}(c) and Fig.~\ref{F3}(a) as the size of the CSRR's split gap increased (from 0.1 to 1.5~$mm$), the peak corresponding to the CSRR mode shifted toward higher frequencies (from 6 to 7.4~$GHz$). The position of the other peak, corresponding to the CELC mode, remained almost constant (6.75~$GHz$) throughout the simulation. As the CSRR mode approached the CELC mode and merged into a unified mode at the coupling centre, resulting in an increase in amplitude. This merging of two modes into a single mode indicated a LA/ CIA phenomenon where the energy levels tend to converge, leading to the merging of the two modes into a single mode. In the stack plots of Fig.~\ref{F2}(c), at the bottom is the lowest 'g' value, where two dips are visible for the same 'g'  one having greater dip  and another having small dip. As 'g' is increasing the small dip vanishes and the big dip is now getting dipper (for 'g'=0.7~mm) with many folds as of previous. Again with further increasing in 'g' value the dept of single unified dip is decreasing and later they are splitted into two dips one greater and one smaller for the same value of 'g'.Here it is obvious that in this system one of the two hybrid modes becomes lossless, and the Friedrich-Wintgen condition is satisfied\cite{friedrich1985interfering,hsu2016bound} with the continuum coupling term $\sqrt{\beta_1 \beta_2}$ near at $\omega_1 - \omega_2 $ near the frequency crossing of uncoupled resonators having $\Delta$=0, thus again satisfying the Friedrich-Wintgen BIC condition numerically as posed by Eq.~\ref{E4}. For this case dissipation of CSRR (CELC) is 0.076 (0.048). In Fig.~\ref{F3}(a), simulation is well consistent with experiments as well as with theory.

Switching to Coupling Induced Transparency (CIT): for case-2 we are rotating only the CSRR Fig.~\ref{F2}(b), we observed two distinct peaks in the frequency spectrum Fig.~\ref{F2}(d) and Fig.~\ref{F3}(b). As we increased the size of the CSRR’s split gap (from 0.1 to 2.2~$mm$), one peak, corresponding to the CSRR mode, shifted towards higher frequencies (from 5.7 to 7.5~ $GHz$). As the CSRR mode peak approached the CELC mode peak (almost constant, 6.65~$GHz$ ), they never coalesced into a single peak. Instead, they exhibited a transparency region at the coupling canter, indicating a distinct level repulsion or Coupling Induced Transparency (CIT) phenomenon. Interestingly, in both cases, as one peak approached the other, there was a gradual change in their amplitudes. The maximum change occurred precisely when one peak crossed over the other, after which they gradually returned to their original magnitudes. For this case the dissipation of CSRR (CELC) is 0.0227 (0.0057) with $\Delta$=0.12.

Fig.~\ref{F4} shows electric field distribution for different regions of interaction of a coupled system, illustrating the coupling mechanism in a hybrid system quite effectively. The distributions are recorded primarily for three regions: far from coupling, at coupling, and near coupling for both cases 1 and 2 respectively. For both cases energy is distributed along both resonators in a coupled system, which shows energy is absorbed by both resonators and showing two peaks in transmission spectra. Near the coupling centre field distribution shows strong coupling. Due to the orientation effect, the coupling centre is different for both cases. For Case 1 at g=0.7~mm (coupling centre) it is seen that most of the EF is concentrated around the CSRR structure, which represents strong absorption at this stage, whereas minimal electric field is seen in CELC. Which is showing that energy at this point is more localised, and not distributed among resonators, which can also be interpreted as one resonance peak verifying emergence of BICs. 
For case-2, at g=1.2mm (coupling centre), opposite to case 1, electric field is distributed among the resonators, which can be understood as sharing of energy between resonators. This energy distribution leads to two resonance peaks at the coupling centre. In case-1, where we see that at coupling centre, energy is more localised, whereas in case-2, distribution of energy is seen at the coupling centre. These two situations govern two phenomena in hybrid a system. Case-1 is interpreted as BIC emergence out of CIA, at coupling centres showing single energy distribution. Case-2 can be interpreted as CIT, a level repulsion case, in which two resonance peaks approach each other; despite merging they start to repel each other, showing two different distributions of energy. 

In summary, for case 1, energy is distributed among the resonators for the far coupling region but becomes localised at the coupling centre where the smaller mode disappeared and larger mode gets many fold for Friedrich-Wintgen BICs. Near the coupling region (for large values of split gap), coupling between resonators starts to weaken. For case-1, intensity of EF is more in CSRR than CELC. This reflects that peak due to CSRR is deeper than CELC, or energy absorbed by CSRR is more as compared to CELC. 
For Case-2(g=1.2mm), at this stage sharing of energy lasts between them, but less dense EF is seen than that at the coupling centre. Which shows, resonators start to seek to their individual nature. EF distribution is seen as more intense in CELC, this reflects that resonance peak due to this is deeper than CSRR or can say more energy is absorbed by CELC. Far From coupling Region (g=2.2~mm), resonance peaks due to two resonators are far apart, showing an individuality of resonators.

\section{Conclusion}
We have fabricated a planar device to explore some peculiar quantum phenomenon of BIC emerging out of CIA as well as CIT, have achieved transition from BIC to CIA in the proposed device by just rotating one of its components with dynamic adjusting of some of its parameters at room temperature. We have shown that in the domain of dissipative coupling cooperative damping can give rise to BICs in the microwave region and may be switched to CIT and vv. Devices considered in this paper are examples of open quantum systems, performing detailed simulation, experiment and theory are in good agreement. Our work for constructing planar devices for holding BIC, CIA, CIT with switching capability holds promises for advancing the developments of open quantum system based devices for future application in suitable quantum information processing.

\begin{acknowledgments}
The work was supported by the Council of Science and Technology, Uttar Pradesh (CSTUP), India, (Project Id: 2470, sanction No: CST/D-1520 and Project Id: 4482, sanction no: CST/D- 7/8). B Bhoi acknowledges support by the Science and Engineering Research Board (SERB) India- SRG/2023/001355
\end{acknowledgments}

\section*{Data Availability Statement}

All data that support the findings of this study are included
within the article (and any supplementary files).

\nocite{*}
\providecommand{\noopsort}[1]{}\providecommand{\singleletter}[1]{#1}%

\end{document}